\documentclass[pra,aps,twocolumn,showkeys,epsfig]{revtex4}
\usepackage{amsmath}
\usepackage{amssymb}
\usepackage{graphicx}
\usepackage{dcolumn}
\usepackage{bm}
\usepackage[colorlinks=false,dvipdfm]{hyperref} 
\usepackage{setspace}
\usepackage{amsfonts}
\usepackage{rotating}
\usepackage{makeidx}
\usepackage{amsthm}

\newcommand{\tr}{\mbox{Tr}}

\begin{document}

\title{Nonequilibrium quantum absorption refrigerator  \footnote{New J. Phys. 20 (2018) 063005}}

\author{Jian-Ying Du}
\affiliation{Department of Physics, School of Science, Tianjin
University, Tianjin 300072, China}

\author{Fu-Lin Zhang}
\email[Corresponding author: ]{flzhang@tju.edu.cn}
\affiliation{Department of Physics, School
of Science, Tianjin University, Tianjin 300072, China}

\date{\today}

\begin{abstract}
We study a quantum absorption refrigerator, in which a target qubit is cooled by two machine qubits in a nonequilibrium steady state.
It is realized by a strong internal coupling in the two-qubit fridge and a vanishing tripartite interaction among the whole system.
The coherence of  a \emph{machine virtual qubit}  is investigated as quantumness of the fridge.
A necessary condition for cooling shows that the quantum coherence is beneficial to the nonequilibrium fridge, while it is detrimental as far as the maximum coefficient of performance (COP) and the COP at maximum power are concerned.
Here, the  COP is defined only in terms of heat currents caused by the tripartite interaction, with the one maintaining the two-qubit nonequilibrium state being excluded.
The later can be considered to have no direct involvement in extracting heat from the target, as it is not affected by the tripartite interaction.
\end{abstract}



\maketitle

\section{Introduction}

Quantum thermodynamics  \cite{Book2004,Book2009}, which investigates the intersection of quantum mechanics and thermodynamics, can be traced back to the early years of quantum mechanics \cite{EPJH2010_1929}.
Significant advances have been made in this  field recently,  especially in the area of out-of-equilibrium thermodynamics \cite{RMP2009nonequilibrium,PRL2011nonequilibrium,EPL2014efficiency,PRA2015quantum} and the interplay with quantum information \cite{PRL2006Canonical,NP2006entanglement,JPA2016}.
The study of quantum thermal machines plays a vital role in understanding the emergence of basic thermodynamic principles at the quantum mechanical level \cite{PRL1959three,PR1967quantum} and uncovering the quantum effects of finite size systems in thermodynamics \cite{S2003extracting,PRL2004second,EPJD2006engines,EPL2014efficiency,PRA2015quantum}.

Recently, the study of small self-contained quantum thermal machines has gained widely attention \cite{ARPC2014quantum,JPA2016}.
The term \textit{small} means few quantum levels, and \textit{self-contained}  (or \textit{autonomous}) refers to the fact that the external control is replaced by their interactions with heat baths at different temperatures.
These machines are also referred to as \textit{continuous} engines, in contrast to the discrete ones with four or two strokes \cite{ARPC2014quantum,PRX2015equivalence}.
Among these, a model of quantum absorption refrigerator \cite{PRL2012quantum} consisting of two machine qubits and a target one has raised a subsequent stream of works \cite{JPA2011smallest,PRE2012virtual,PRE2013performance,PRE2014re,PRE2014entanglement,SR2014quantum}, since it was proposed in the investigation of the fundamental limitation on the size of thermal machines \cite{PRL2010small}.

A fundamental topic in these researches is to establish the role of quantumness.
Recent work \cite{SR2014quantum} shows the advantages of  quantum properties represented by spectral structure of thermal reservoirs.
The role of quantum features in the models are investigated in the both regimes of weak \cite{PRE2014entanglement,PRA2015quantum} and strong \cite{PRE2013performance} coupling.
In these studies, the quantum features refer to quantum correlations, measured by entanglement \cite{RMP2009Ent} and discord \cite{RMP2012Discord}, which originate from the global coherence \cite{PRL2014Coherence} among the whole system of machine and target.
In addition, the continuous heat machines are shown to operate on coherence \cite{PRX2015equivalence}.
However, the internal quantumness among different particles in a machine, e.g. the coherence between the spiral and engine in the three-qubit model, is not involved in these works.
One can consider the quantum correlations in \cite{PRE2013performance,PRE2014entanglement,PRA2015quantum} as the internal quantum properties of machines, by regarding the target qubit as part of the refrigerator, and its bath as the object to cool.
But this view pollutes the simplicity of the model.

In this paper, we address this problem by introducing a strong internal coupling between the engine and spiral in the three-qubit model of absorption refrigerator, while the tripartite interaction extracting energy from the target is supposed to be weak enough as in the original construction \cite{PRL2010small}.
The former generates the quantum coherence in the fridge and the later allows us to talk about the local temperature \cite{PRL2010small,PRX2014locality,PRL2007theory} of the target, and hence its cooling.
When the three-body interaction is turned off,  the two machine qubits are in a  nonequilibrium stationary state with a global coherence between them, and  the target qubit is in a thermal state at its bath temperature.
We term the machine as \textit{nonequilibrium fridge}, since the task of cooling is mainly dependent upon the nonequilibrium state and its thermal contact with the target via the arbitrarily weak tripartite interaction \cite{PRE2012virtual}.




We choose the subspace of fridge with two-qubit coherence as the \textit{machine virtual qubit} \cite{PRE2012virtual} acting on the target directly, and adopt its quantum coherence as a measure of the quantumness involved in to the task of cooling.
As our main result, such coherence is shown to be beneficial to the fridge by a necessary condition for cooling; but to be a disadvantage by COP, although the heat current maintaining the coherence is excluded in our definition of COP.
This disadvantage is not natural like the one of tripartite coherence in \cite{PRE2014entanglement}, which is positively correlated with the heat currents driven by temperature differences and thus is a reflection of irreversibility.

Our treatment of the master equation is a hybrid of the global and local approaches \cite{EPL2014local,EPL2016perturbative}.
Namely, the two-qubit fridge is considered as a whole.
Dissipations will not destroy its eigenstates but only produce transitions between them.
Whereas, the tripartite interaction is assumed to be too weak to affect the system-bath coupling.
In addition to the study of quantumness in quantum machines, our construction can serve as a simpler model to understand the effects of delocalized dissipations on the task of cooling in \cite{PRE2013performance,PRE2014re}.

In the next section, we introduce our model and derive the master equation.
Based on the stationary state, the role of the coherence of machine virtual qubit is analyzed in Sec. \ref{Coherence}, by studying a necessary condition for cooling, the maximal COP, and the COP at maximum cooling power.
Sec. \ref{Conclu} presents the conclusions.

\section{Fridge with internal interaction}\label{Model}

\subsection{Model}

\begin{figure}
\includegraphics[width=6cm]{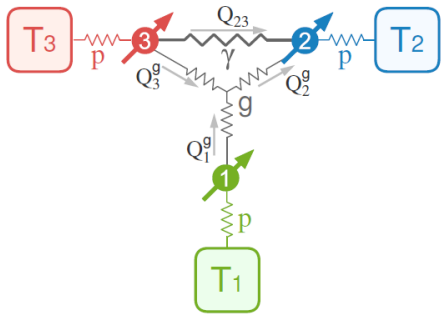}
\caption{
(Color online) Schematic diagram of the quantum refrigerator.
Three qubits couple their respective baths at temperatures $T_1 \leq T_2 \leq T_3$  with a strength $p$.
The qubits $1,2,3$ are the target, spiral and engine, in order.
The two-qubit interaction with strength $\gamma$ between the spiral and engine builds their coherence and generates a heat current $Q_{23}$.
Heat currents $Q_{1}^g$, $Q_{2}^g$, $Q_{3}^g$ are caused by the tripartite interaction $g$.
 }
\label{FigModel}
\end{figure}

The model we consider here is made up of three qubits, $1$, $2$, and $3$, which interact with three baths at temperatures $T_1<T_2<T_3$ respectively (see Fig. \ref{FigModel}).
Qubit $1$ is the \textit{target} to be cooled,  while the other two play the role of a fridge, in which qubit $2$ is the \textit{spiral} that extracts heat from the target and qubit $3$ is the \textit{engine} providing free energy.
The free Hamiltonian for the three qubits is
\begin{equation}
\mathcal{H}_{0}= \mathcal{H}_{1}+ \mathcal{H}_{2}+ \mathcal{H}_{3},
\end{equation}
where $\mathcal{H}_{i}=E_{i}\sigma_i^{z}/2$, $i=1,2,3$, with $\sigma_i^{z}=|0\rangle\langle0|-|1\rangle\langle1| $ being the third Pauli operator.
Without the interactions among qubits,  each qubit is in a thermal state as $\tau_i=r_i|0\rangle\langle0|+\bar{r}_i|1\rangle\langle1| $, where $r_{i}=1/[1+\exp(\beta_i E_{i})]$ ,  $\bar{r}_{i}=1-r_{i}$ and $\beta_i= 1/T_i$.

To biuld coherence in the fridge we introduce a two-qubit interaction
\begin{equation}
 \mathcal{H}_{\gamma}= \gamma (\sigma_2^{+} \sigma_3^{-}+\sigma_2^{-} \sigma_3^{+}),
\end{equation}
where $\sigma_i^{+}=|0\rangle\langle1|$,  $ \sigma_i^{-}=|1\rangle\langle0|$ and $\gamma$ is the interaction strength.
It is assumed  to be comparable with the free Hamiltonian ($\gamma \sim E_i$), and much larger than the qubit-bath couplings.
That is, the dissipations will not destroy the eigenstates of the fridge  governed by Hamiltonian $\mathcal{H}_{\rm fridge}=\mathcal{H}_{2}+\mathcal{H}_{3}+ \mathcal{H}_{\gamma}$.
Consequently, the nonequilibrium steady state of fridge is a  mixture of its eigenstates.

We choose the two eigenstates of  $\mathcal{H}_{\rm fridge}$ with two-qubit coherence as the \textit{machine virtual qubit} $v$, and ensure its energy-level spacing equal to $E_1$ by adjusting the values of $E_2$, $E_3$ and $\gamma$.
The task of cooling the target can be performed by an arbitrarily weak interaction which allows the virtual qubit and target to resonantly exchange energy, which is given by
\begin{eqnarray}\label{TriH}
 \mathcal{H}_{g}= g(\sigma_1^{+}\sigma_v^{-}+ \sigma_1^{-}\sigma_v^{+}),
\end{eqnarray}
 where $\sigma_v^{\pm}$ are the raising  and lowering operators for the virtual qubit.
This interaction places the target and fridge, via virtual qubit $v$, in thermal contact \cite{PRE2012virtual}.

\subsection{Master equation}

The master equation governing the dynamics of the three-qubit system is given by
 \begin{equation}\label{MasterEq}
 \dot{\rho} = -i[\mathcal{H}_1+\mathcal{H}_{\rm fridge},\rho]-i[\mathcal{H}_g,\rho]+ \sum_{i=1}^{3}\mathcal{D}_{i}(\rho),
\end{equation}
where $\mathcal{D}_{i}$ are  the dissipative completely positive superoperators for each bath.
In the weak-tripartite-coupling regime, $g \ll E_i$, the eigenvalues and eigenstates of the three-qubit system remain governed by the Hamiltonian $\mathcal{H}_1+\mathcal{H}_{\rm fridge}$, and hence we can meaningfully talk about the temperature of the target qubit, since it will remain in the standard thermal state with the energy gap $E_1$ \cite{PRL2010small}.

We model the local dissipator for each bath on its corresponding free qubit (when $g=0$ and $\gamma=0$)  as
\begin{eqnarray}\label{Dlocal}
   \mathcal{D}^{f}_{i}(\rho)=p_{i} \biggr[ r_{i}\bigr(\sigma_i^{+}\rho\sigma_i^{-}- \frac{1}{2} \{ \sigma_i^{-}\sigma_i^{+}, \rho\} \bigr) \ \ \  \ \ \ &&  \nonumber  \\
           +    \bar{r}_{i} \bigr( \sigma_i^{-}\rho\sigma_i^{+}- \frac{1}{2} \{ \sigma_i^{+}\sigma_i^{-}, \rho\} \bigr) \biggr], &&
\end{eqnarray}
where $p_i$  is the dissipation rate.
It is a variant of the simple reset model in \cite{PRL2010small,JPA2011smallest,PRE2014entanglement} with the dephasing rate being halved.
In this work, we suppose $p_1=p_2=p_3=p$ for simplicity.
In the regime where $p \sim g \ll E_i$, the dissipator $\mathcal{D}_{1}=\mathcal{D}^{f}_{1}$ acts locally on the target, while the other two on  the whole of the two-qubit fridge due to the internal coupling $H_{\gamma}$.

The form of $\mathcal{D}^{f}_{i}$ in (\ref{Dlocal}) can be derived by using a specific system-environment model, such as the atom-field system with an appropriate spectral density under the Markovian and rotating-wave approximations \cite{BookOpenSys,Book2007QO}.
Sequentially, one can take into account the two-qubit coupling $H_{\gamma}$ and obtain the delocalized dissipators, $\mathcal{D}_{2}$ and $\mathcal{D}_{3}$, in the same model, as the processes in \cite{PRE2013performance,PRE2014re,ACMP2015nonequilibrium,OC2017two}.
The procedures are equivalent to the transformation, from the transitions between the eigenstates of free qubits into the ones of $\mathcal{H}_{\rm fridge}$, in the following two steps.




Firstly, one can diagonalize the fridge Hamiltonian into two free qubits by using the unitary
\begin{equation}\label{UDiag}
\mathcal{U}=\cos^2\frac{\theta}{4}+\sin^2\frac{\theta}{4}\sigma^z_2 \sigma^z_3 +\sin\frac{\theta}{2}(\sigma_2^{+}\sigma_3^{-}-\sigma_2^{-}\sigma_3^{+}),
\end{equation}
where  $\theta=\arctan (2\gamma/\Delta E)$ with $\Delta E=E_2-E_3$.
They satisfy
\begin{equation}\label{Diag}
\mathcal{H}_{\rm fridge}=\mathcal{U}^{\dag}\biggr(\varepsilon_{2}\frac{\sigma_2^{z}}{2}+\varepsilon_{3}\frac{\sigma_3^{z}}{2}\biggr)\mathcal{U},
\end{equation}
where $\varepsilon_{2}= E + \lambda $, $\varepsilon_{3}= E - \lambda$, with $E=(E_2+E_3)/2$ and $\lambda=\sqrt{(\Delta E/2)^2+\gamma^2}$.
 In the following, we denote the two free qubits in the diagonalized fridge as $\tilde{\mu}=\tilde{2},\tilde{3}$, and their Pauli operators $\tilde{\sigma}^{\pm,z}_{\mu} =\mathcal{U}^{\dag} \sigma^{\pm,z}_{\mu} \mathcal{U}$.
The four eigenvalues of the fridge are $\{E,\lambda,-\lambda,-E\}$ and their corresponding eigenvectors  are
$|\psi_{00}\rangle=\mathcal{U}^{\dag} |00\rangle$, $|\psi_{01}\rangle=\mathcal{U}^{\dag} |01\rangle$, $|\psi_{10}\rangle=\mathcal{U}^{\dag} |10\rangle$, $|\psi_{11}\rangle=\mathcal{U}^{\dag} |11\rangle$.
Here,  $|\psi_{01}\rangle$ and $|\psi_{10}\rangle$ are two coherent superpositions of states $|01\rangle$ and $|10\rangle$, while $|\psi_{00}\rangle= |00\rangle$ and $|\psi_{11}\rangle= |11\rangle$ are two direct product states.
The machine virtual qubit is defined as the subspace $\{ |\psi_{01}\rangle, |\psi_{10}\rangle  \}$, and the  raising  and lowering operators in Eq. (\ref{TriH})
 are $\sigma^{+}_v=|\psi_{01}\rangle  \langle \psi_{10}|=\tilde{\sigma}_2^+ \tilde{\sigma}_3^-$ and $\sigma^{-}_v= |\psi_{10}\rangle  \langle \psi_{01}|=\tilde{\sigma}_2^- \tilde{\sigma}_3^+$.
 The resonant interaction  requires that $E_1=\varepsilon_{2}-\varepsilon_{3}=2 \lambda$.

Secondly, to protect the eigenstates of $\mathcal{H}_{\rm fridge}$,   we shall map $\sigma^{\pm}_{\mu=2,3}$ in Eq. (\ref{Dlocal}) onto the jump operators
 $\Gamma^{\pm}_{\nu \mu}= \sum'_{E_{ij}-E_{kl}= \omega_\nu} |\psi_{ij}\rangle  \langle \psi_{ij}|  \sigma^{\pm}_{\mu}   |\psi_{kl}\rangle  \langle \psi_{kl}|$,
  with  $\omega_\nu$ being the energy-level spacings.
There are four pairs of nonzero $\Gamma^{\pm}_{\nu \mu}$ as  $\Gamma^{\pm}_{2 2}=\cos \frac{\theta}{2}  \tilde{\sigma}^{\pm}_2 $,  $\Gamma^{\pm}_{3 2}=\sin \frac{\theta}{2}  \tilde{\sigma}^z_2 \tilde{\sigma}^{\pm}_3 $, $\Gamma^{\pm}_{33}=\cos \frac{\theta}{2}  \tilde{\sigma}^{\pm}_3 $, and $\Gamma^{\pm}_{2 3}=-\sin \frac{\theta}{2}  \tilde{\sigma}^{\pm}_2 \tilde{\sigma}^z_3 $,
 corresponding to $\omega_{\nu}=\pm\varepsilon_{\nu}$.
 Let $r_{\nu\mu}=1/[1+\exp(-\beta_\mu \varepsilon_{\nu})]$ and  $\bar{r}_{\nu\mu}=1-r_{\nu\mu}$ with $\mu,\nu=2,3$, which are the probabilities for a two-level thermal state in the  temperature $T_{\mu}$, with energy-level spacing $\varepsilon_{\nu}$, in the ground and excited state respectively.
Replacing $\sigma^{\pm}_{\mu}$ and $r_{\mu}$ in the dissipator $\mathcal{D}^f_{\mu}$ in Eq. (\ref{Dlocal}) with by $\Gamma^{\pm}_{\nu \mu}$ and $r_{\nu\mu}$, one obtains four delocalized dissipators, denoted by $\mathcal{D}_{\nu \mu}$.
Then, the dissipators, describing the effects of baths $2$ and $3$, are given by $\mathcal{D}_{\mu=2,3}=\mathcal{D}_{2 \mu}+\mathcal{D}_{3 \mu}$.

%

\section{Cooling with coherence}\label{Coherence}

\subsection{Stationary state}

\begin{figure}
\includegraphics[width=6cm]{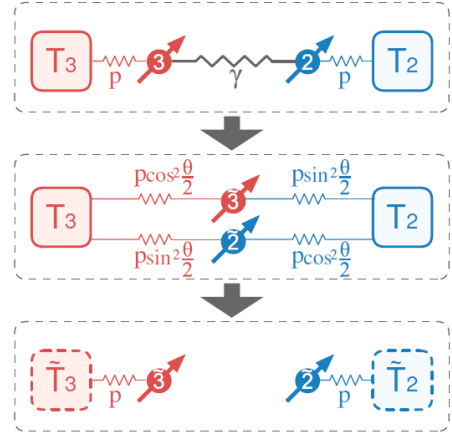}
\caption{(Color online)
The process to localize the channels $\mathcal{D}_{\mu=2,3}$.
The fridge Hamiltonian can be diagonalized into two free qubits $\tilde{2}$ and $\tilde{3}$, both of which couples with the two baths $2$ and $3$.
The effect of the two baths on a state $\tilde{\rho}$ in Eq. (\ref{Sstate}) is equivalent to the thermalization of qubit $\tilde{\nu}$ under the influence of bath $\mu$ with dissipation rates ${p}\cos^2 \frac{\theta}{2}$ or ${p}\sin^2 \frac{\theta}{2}$.
The net effect is that baths $2$ and $3$ draw the qubit $\tilde{\nu}$ back to a thermal state in $\tilde{T}_{\nu}$ with a rate ${p}$.
} \label{FigModelTrans}
\end{figure}

We are interested in the steady-state solution of  the master equation (\ref{MasterEq}), satisfying $\dot{\rho}_S=0$.
It can be derived simply by \textit{localizing} the channels $\mathcal{D}_{\mu=2,3}$ in the representation of diagonalized fidge.
The process to localize the channels is shown in Fig. \ref{FigModelTrans}.

Specifically,  we define four local dissipators on the  free qubits in the diagonalized fridge as,
$\tilde{\mathcal{D}}_{2 2}= \mathcal{D}_{22}$, $\tilde{\mathcal{D}}_{23}= \mathcal{D}_{23}|_{\Gamma^{\pm}_{2 3} \to\sin \frac{\theta}{2}  \tilde{\sigma}^{\pm}_2 }$, $\tilde{\mathcal{D}}_{32}= \mathcal{D}_{32}|_{\Gamma^{\pm}_{3 2}\rightarrow\sin \frac{\theta}{2}  \tilde{\sigma}^{\pm}_3 }$,  and $\tilde{\mathcal{D}}_{33}= \mathcal{D}_{33}$.
One can consider a class of states in the form as
\begin{eqnarray}\label{Sstate}
\tilde{\rho}=\frac{1}{8}\biggr(1+\sum_{i} a_i \tilde{\sigma}_i^z  +\sum_{i \neq j} b_{ij} \tilde{\sigma}_i^z\tilde{\sigma}_j^z + c\tilde{\sigma}_1^z\tilde{\sigma}_2^z \tilde{\sigma}_3^z  + d \mathcal{Y}\biggr ),
\end{eqnarray}
 where $\mathcal{Y}=-i \tilde{\sigma}_1^+ \tilde{\sigma}_2^- \tilde{\sigma}_3^+ + i \tilde{\sigma}_1^- \tilde{\sigma}_2^+ \tilde{\sigma}_3^-$ and $\tilde{\sigma}_1^{\pm,z}=\sigma_1^{\pm,z}$.
The effect of $\mathcal{D}_{2,3}$  on these states is equivalent to  two local dissipators that
\begin{eqnarray}
\mathcal{D}_{2}(\tilde{\rho})+\mathcal{D}_{3}(\tilde{\rho}) = \tilde{\mathcal{D}}_{2}(\tilde{\rho})+\tilde{\mathcal{D}}_{3}(\tilde{\rho}),
\end{eqnarray}
with $\tilde{\mathcal{D}}_{\nu}=\tilde{\mathcal{D}}_{\nu 2}+\tilde{\mathcal{D}}_{\nu  3 }$.
Here, $\tilde{\mathcal{D}}_{\nu}$ are in the form of (\ref{Dlocal}) with the Pauli operators of qubit $\tilde{\nu}$,  dissipation rates $\tilde{p}_{\nu}=p$, and  the probabilities $\tilde{r}_{\nu}=  \cos^2 \frac{\theta}{2}r_{\nu \nu }+ \sin^2 \frac{\theta}{2}r_{\nu \mu} $ and $\bar{\tilde{r}}_{\nu}=  1-\tilde{r}_{\nu} $.
That is, the delocalized dissipations thermalize the qubit $\tilde{\nu}$ to a thermal state $\tilde{\tau}_{\nu}=\tilde{r}_{\nu}|0\rangle \langle0|+\bar{\tilde{r}}_{\nu}  |1\rangle \langle1|$ in a temperature $\tilde{T}_{\nu}= \varepsilon_{\nu}/[ \ln(\bar{\tilde{r}}_{\nu} /\tilde{r}_{\nu})]$.

A localized master equation can be obtained by replacing $\mathcal{D}_{2,3}$  in (\ref{MasterEq}) with $\tilde{\mathcal{D}}_{2,3}$.
 Its stationary-state solution is in the form as (\ref{Sstate}), and therefore is also the stationary state of the master equation  (\ref{MasterEq}),  $\rho_S$.
Without the tripartite interaction, the total stationary state is simply the direct product of three thermal state $\rho_{S,0}=\tau_1 \tilde{\tau}_{2} \tilde{\tau}_{3}$.
The tripartite interaction generates a deviation from $\rho_{S,0}$ proportional to the parameter
 \begin{equation}\label{d}
d= \frac{48 ( \bar{r}_1 \tilde{r}_2  \bar{\tilde{r}}_3-r_1 \bar{\tilde{r}}_2 \tilde{r}_3 )}{9 p^2+ (14 +4\sum_{i \neq j} \Omega_{ij}) g^2} pg,
\end{equation}
where $\Omega_{12}=r_1 \bar{\tilde{r}}_2+\bar{r}_1 \tilde{r}_2$,  $\Omega_{23}=\tilde{r}_2 \bar{\tilde{r}}_3+\bar{\tilde{r}}_2 \tilde{r}_3$, and $\Omega_{31}=r_1 \tilde{r}_3+\bar{r}_1 \bar{\tilde{r}}_3$.
And then the other parameters for the stationary state are $a_{1,3}=s_{1,3}+\frac{g}{p}\frac{d}{2}$, $a_{2}=s_{2}-\frac{g}{p}\frac{d}{2}$, $b_{ij}=\frac{1}{2}(s_i a_j+s_j a_i)$, and $c=\frac{1}{3}(\sum_{i\neq j\neq k}   s_i b_{jk} -\frac{g}{p}\frac{d}{2})$.
Here $s_i$ stand for the Bloch vectors, $s_1=r_1-\bar{r}_1$ and $s_{2,3}=\tilde{r}_{2,3}-\bar{\tilde{r}}_{2,3}$.

One can notice that, the total stationary state $\rho_S$ is determined by the properties of $\rho_{S,0}$ and the coupling strengths  $p$ and $g$.
It is the result of competition between the trend back to $\rho_{S,0}$  and the thermal contact of fridge with the target.
In this sense, the task of cooling can be regarded as that, the target qubit is cooled by the fridge in a nonequilibrium steady-state $\rho_{\rm fridge} = \tilde{\tau}_{2} \tilde{\tau}_{3}$ in our model, while in the direct product of two thermal states in the original construction \cite{PRL2010small}.

In the nonequilibrium fridge, the virtual machine qubit is the part acting on the target directly.
Its \textit{virtual temperature}  can be found by looking at the ratio of populations of $|\psi_{01}\rangle$ and $|\psi_{10}\rangle$ in $\rho_{\rm fridge} $,
\begin{equation}\label{Tv}
T_v= \frac{\varepsilon_2-\varepsilon_3}{ \ln [(\bar{\tilde{r}}_2 \tilde{r}_3)/(\tilde{r}_2 \bar{\tilde{r}}_3)]} =  \frac{\varepsilon_2-\varepsilon_3}{ \varepsilon_2/\tilde{T}_2  -\varepsilon_3/\tilde{T}_3}.
\end{equation}
The coherence of the virtual qubit can be measured by its  nondiagonal elements,
corresponding to the coherent superpositions of $|01\rangle$ and $|10\rangle$, as \cite{PRL2014Coherence}
\begin{equation}
C(\rho_v)=\biggr| \frac{\tilde{r}_2-\tilde{r}_3}{\tilde{r}_2+\tilde{r}_3-2\tilde{r}_2 \tilde{r}_3} \biggr| \sin \theta.
\end{equation}

\subsection{Heat currents and COP}

To quantify the performance of the fridge, we derive the stationary heat currents in the model.
The ones flowing from the three baths are defined as $Q_i=\tr [\mathcal{H}_{\rm tot}\mathcal{D}_i(\rho_S)]$ \cite{BookOpenSys}, where the total Hamiltonian $\mathcal{H}_{\rm tot}=\mathcal{H}_1+\mathcal{H}_{\rm fridge}+\mathcal{H}_{g}$.
They are given by
\begin{eqnarray}\label{Qi}
&&Q_1=-\frac{1}{4} gd E_1, \nonumber \\
&&Q_2=-Q_{23}+ \frac{1}{4} gd (\varepsilon_2 \cos^2 \frac{\theta}{2} - \varepsilon_3 \sin^2\frac{\theta}{2}),\\
&&Q_3=Q_{23}- \frac{1}{4} gd (\varepsilon_3 \cos^2 \frac{\theta}{2} - \varepsilon_2 \sin^2\frac{\theta}{2}), \nonumber
\end{eqnarray}
with $Q_{23}=\tr [\mathcal{H}_{\rm fridge}\mathcal{D}_3(\tilde{\tau}_{2} \tilde{\tau}_{3})]=-\tr [\mathcal{H}_{\rm fridge}\mathcal{D}_2(\tilde{\tau}_{2} \tilde{\tau}_{3})]$.

The assumption of weak interaction allows us to define the heat flow drawn from the target by the fridge as $Q^g_1=- \tr [ \mathcal{H}_1 \mathcal{D}_g(\rho_S) ]$ with $\mathcal{D}_g(\rho_S)=- i[ \mathcal{H}_{g},\rho_S]$, which is the change in local energy due to the tripartite interaction.
It is straightforward to check the conservation of energy that $Q^g_1=Q_1$.
The current injected by the hot bath, $Q_3$ in Eq. (\ref{Qi}), consists of two parts, of which $Q_{23}$ is independent of the tripartite interaction, while the rest $Q^g_3=Q_3-Q_{23}$ is proportional to the product of interaction ($g$) and deviation of $\rho_{S}$ from $\rho_{S,0}$ ($d$).
The former can be considered to have no direct involvement in the task of cooling, but plays a role to maintain the nonequilibrium state $\rho_{\rm fridge}$ and hence the coherence of virtual qubit.
Accordingly, the later is the extra free energy gained by the nonequilibrium state $\rho_{\rm fridge}$ from the hot bath to cool the target.

From this point of view, we define a \textit{COP} of our nonequilibrium fridge as
 \begin{equation}\label{etag}
\eta_g=\frac{Q_1^g}{Q_3^g}=\frac{E_1}{\varepsilon_3 \cos^2 \frac{\theta}{2} - \varepsilon_2 \sin^2\frac{\theta}{2}}.
\end{equation}
It is independent of the deviation of the total stationary state from $\rho_{S,0}$, and thus can be used to quantify the performance of the nonequilibrium state $\rho_{\rm fridge}$ in cooling the target in $\tau_1$.
We will compare it to the thermodynamic COP  $\eta_{  tot}= {Q_1}/{Q_3}$, that is, for a given supply of energy from the hot bath, how much heat can be extracted from the target  \cite{JPA2011smallest}.
In this article, references to COP are to the one in Eq. (\ref{etag}), unless otherwise stated.


\subsection{Necessary condition for cooling}

In this and subsequent parts, we give some analysis based on a combination of analytical and numerical results.
The reduced state of the target, $\rho_1= \tr_{23} \rho_S$, is in the standard thermal form with the Bloch vector $a_1$.
Its local temperature, $T_1^S$,  is defined by the ratio of populations of ground and excited states \cite{PRE2012virtual}.
To testify the rationality of this local temperature, one can couple the target with a bath in $T_1^S$, whose effect is described by a local dissipator, $\mathcal{D}_1^S$, in the form (\ref{Dlocal}), and find that the heat flow $ \tr [\mathcal{H}_{\rm tot}\mathcal{D}_1^S(\rho_S)]=0$.

Cooling of the target means reaching a temperature $T_1^S<T_1$, corresponding to a negative $d$, and  hence a positive $Q_1^g$.
 It is equivalent to that the numerator in Eq. (\ref{d}) is negative, and consequently $T_1>T_v$.
The energy flow $Q_1^g$ can be taken as a sign of cooling.
We would like begin with the spacial case with $T_1=T_2$, which is the easiest case to achieve cooling.
For this two-qubit machine to act as a fridge, the heat current should satisfies $Q_1^g>0$, under the temperatures $T_1 \leq T_2 \leq T_3$.





%

\begin{figure}
 \begin{flushright}
\includegraphics[width=8.54cm]{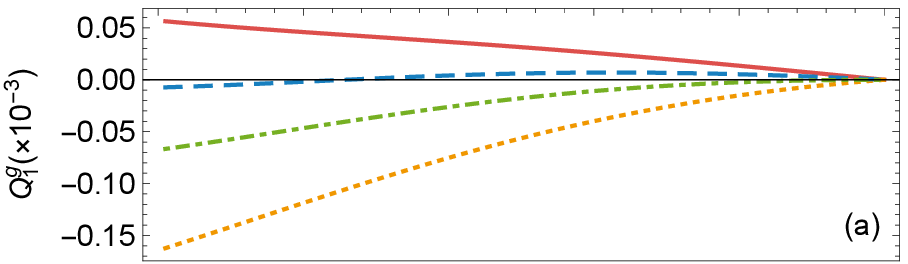}\\
\includegraphics[width=8.50cm]{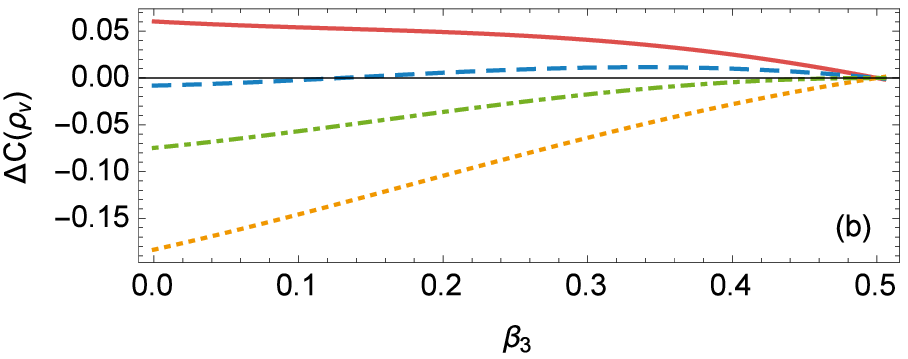}
\end{flushright}
\caption{(Color online)
Heat current $Q_1^g$ in (a) and the change of coherence $\Delta C(\rho_v)$ in (b) as functions of $\beta_3$.
We use the parameters given by $T_1=T_2=2$, $p=g=0.01$, $E_1=1$, $E_3=4$, and from top to bottom $\gamma=0.48,0.49,\sqrt{2\sqrt{17}-8}$ (critical value of the inequality (\ref{Cond})) , $0.50$.
} \label{FigQ1VsB3}
\end{figure}

We keep $E_1$ and $E_3$ invariant and plot the heat currents $Q_1^g$ as functions of $\beta_3$ in Fig. \ref{FigQ1VsB3} (a) at different values of $\gamma$.
It is obvious that the internal interaction between the two machine qubits directly leads to the suppression of $Q_1^g$.
As the interaction increases, there exist three types of curves in turn:
(1) $Q_1^g$ increases with $T_3$;
(2) $Q_1^g$ increases first and then decreases, and when bath $3$ is hot
enough $Q_1^g<0$;
(3) $Q_1^g<0$, and its absolute value increases with $T_3$.
These are very similar with  the results in the study of the three-qubit model with a strong tripartite interaction \cite{PRE2014re},  which is described by a global master equation.

These phenomenons can be  understand  with the aid of  the picture of diagonalized fridge and the virtual temperature.
It is easily to notice that, the virtual temperature in  Eq. (\ref{Tv}) can be reduced by raising $\tilde{T}_3$ or dropping $\tilde{T}_2$.
However, due to the delocalized dissipation effect, both of $\tilde{T}_2$ and $\tilde{T}_3$ are increased, when bath $3$ is heated up  from the temperature $T_2$.
When $\gamma$ or $T_3$ becomes large enough, $T_v>T_2$ and thus the current $Q_1^g<0$.

The two-qubit machine serves as a fridge when it extracts heat from a cold qubit, whose bath is in the temperature  $T_1\leq T_2$.
For the easiest case, $T_1=T_2$, it is required that, the hot bath in $T_3 \geq T_2$ reduces the virtual temperature to less than $T_2$.
That is,  the derivative $\partial T_v/\partial T_3 < 0$ at the point $T_3=T_2$,
leading to
 \begin{equation}\label{Cond}
2 \gamma^2 <  E_3 \Delta E,
\end{equation}
and corresponding to the first two types of curves.
It is a necessary condition for the two-qubit machine to act as a fridge, which does not depend on $T_2$, but is a general requirement on the  two-qubit Hamiltonian.
More fortunate is that it is equivalent to that the denominator in the expression of $\eta_g$ in Eq. (\ref{etag}) is  positive.
In other words, a positive $Q_1^g$ drawn from the target always requires a positive $Q_3^g$ provided by the hot bath.
Consequently, our definition of the COP $\eta_g$ does not violate the second law of thermodynamics at this point.

When the hot bath is heated up from $T_2$, the trend of virtual temperature is determined by the change of populations of the excited and ground state in the virtual qubit, which is accompanied by the change of virtual qubit coherence simultaneously.
In Fig. \ref{FigQ1VsB3} (b), we plot the changes of coherence $\Delta C(\rho_v) =C(\rho_v)-C(\rho_v|_{T_3=T_2})$ as functions of $\beta_3$.
One will immediately see the similarity between $\Delta C(\rho_v)$ and $Q_1^g$.
They have the same zero points, where the state of virtual qubit is unchanged although both of $\tilde{\tau}_2$ and $\tilde{\tau}_3$ are different with the ones in the case of $T_3=T_2$.
In order to understand the similarity, one can consider a given two-qubit system, described by the Hamiltonian $\mathcal{H}_{\rm fridge}$, and thermalized by the contact between qubit $2$ and bath $2$.
The classical probabilities in the equilibrium state reduce the quantum coherence of the virtual qubit.
To make the two-qubit machine to become a fridge, we contact qubit $3$ with a bath $3$ at a hotter temperature $T_3>T_2$.
Only when the thermal contact lowers the virtual temperature  $T_v < T_1$,  the current $Q_1^g>0$ and the target is cooled.
Simultaneously, the coherence $C(\rho_v)$ is enhanced, as it is a monotony decrease function of $T_v$.
When $T_1 < T_2$, the region of positive $Q_1$ becomes a subinterval of the one with $\Delta C(\rho_v)>0$.
Therefore,  the task of cooling requires the the increase of  $ C(\rho_v)$ in the thermal contact with bath $3$.
That is, the virtual qubit coherence is beneficial to the fridge.

\subsection{Maximum COP}

\begin{figure}
 \begin{flushright}
\includegraphics[width=8.37cm]{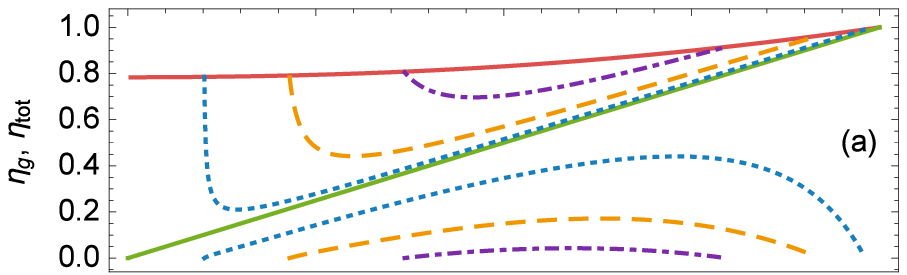}\\
\includegraphics[width=8.31cm]{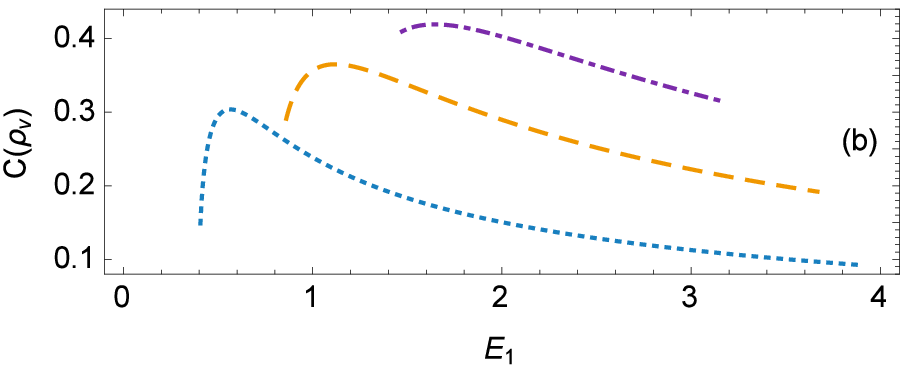}
\end{flushright}
\caption{
(Color online)
COPs $\eta_g$ (above the  solid  green line) and $\eta_{tot}$ (under the solid green  line) in (a), and $C(\rho_v)$ in (b), as functions of $E_1$.
The parameters are $E_3=4$, $T_1=4/3$, $T_2=2$, $T_3=4$, and $\gamma=0.2$ (dotted blue), $0.4$ (dashed yellow), and $0.6$ (dot-dashed purple).
The solid red line shows the upper bound of $\eta_g$, and the green one shows the amount of $E_1/E_3$.
} \label{FigEtagMaxVsE1}
\end{figure}

Now we turn to the COP $\eta_g$ of the nonequilibrium fridge defined in Eq. (\ref{etag}).
We plot the amounts of $\eta_g$, together with $\eta_{tot}$ and $C(\rho_v)$, as functions of $E_1$ in Fig. \ref{FigEtagMaxVsE1} with different values of $\gamma$ in the regions of $Q_1^g>0$, $Q_3^g>0$ and $T_1<T_2<T_3$, where the two machine qubits act as a fridge.

\begin{figure}
 \begin{flushright}
\includegraphics[width=8.41cm]{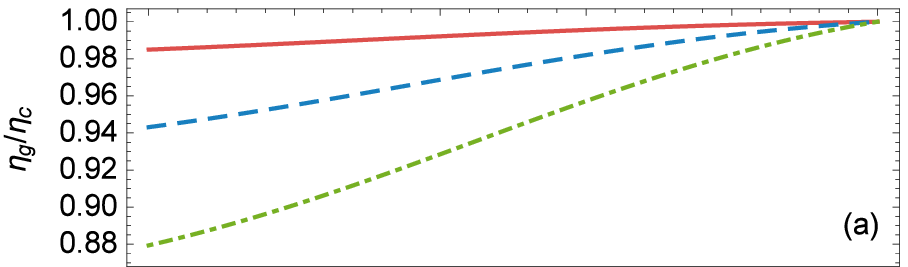}\\
\includegraphics[width=8.17cm]{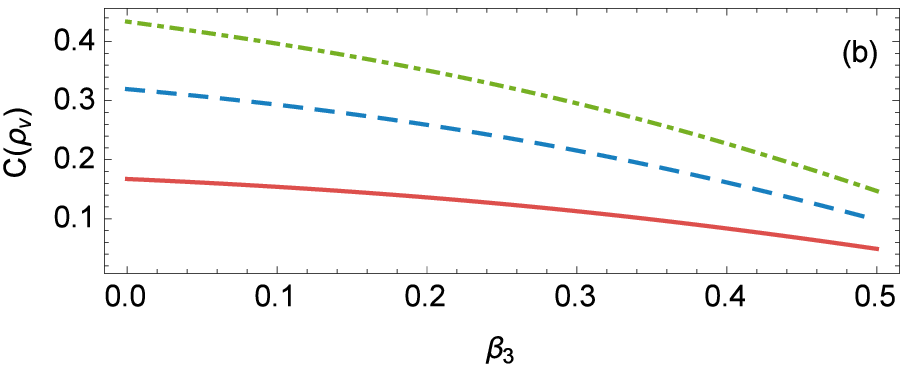}
\end{flushright}
\caption{(Color online)
The upper bound of $\eta_g$ normalized by $\eta_c$ in (a) and the amount of coherence of the virtual qubit $C(\rho_v)$ in (b) as functions of $\beta_3$, for fixed $E_1=1$, $E_3=4$, $T_2=2$, and $\gamma=0.1$ (solid red), $0.2$ (dashed blue), and $0.3$ (dot-dashed green).
} \label{FigEtagMaxVsB3}
\end{figure}

Obviously,  the COP  $\eta_g$ is enhanced by the internal coupling of fridge, while the price is that the region of cooling is reduced.
The two endpoints of the region of cooling are two solutions of $T_1=T_v$, corresponding to the deviation $d = 0$.
These endpoints form the upper bound of $\eta_g$  for fixed $E_1$ and $T_{i=1,2,3}$, which is lower than the Carnot performance.
There are two cases where the upper bound saturates $\eta_c$: (1) the right endpoint with $\gamma\to0$ as shown in Fig. \ref{FigEtagMaxVsE1};
(2) the temperature  $T_3 \to T_2$ as shown in Fig. \ref{FigEtagMaxVsB3} (a).

In contrast, the thermodynamic COP $\eta_{tot}$  is reduced by the two-qubit interaction from the amount of $E_1/E_3$, which corresponds to the case with $\gamma=0$.
And it vanishes when $E_1$ approaches the endpoints of the cooling windows, where $Q_1^g$ becomes zero.
These two properties come mostly from the internal heat current of the fridge, $Q_{23}$, which increases with $\gamma$ and is greater than zero at the endpoints of cooling region.
 For brevity, we omit the concrete form of  $Q_{23}$.
The global coherence and $Q_{23}$ are two aspects of the integrity of the two-qubit fridge.
It can be expected that, the COP $\eta_g$, as the \textit{rest} of $\eta_{tot}$ after the removal of $Q_{23}$, can better reflect the effect of quantum coherence in cooling.


It is shown in Fig. \ref{FigEtagMaxVsE1} that, the coherence $C(\rho_v)$ exhibits the opposite behaviors of $\eta_g$ for a fixed $\gamma$.
As $E_1$ decreases, $C(\rho_v)$ increases to a maximum and then decrease sharply, while  $\eta_g$ experiences a decrease and a sharp increase.
One can understand such phenomena in the viewpoint of virtual temperature.
For a large $E_1$, the parameter $\theta$ approaches $0$, the delocalized dissipation effects in the fridge are small.
The virtual qubit can be cooled by decreasing $E_1$, and thereby COP is reduced, as the case without the two-qubit coupling \cite{PRL2010small}.
However, as $E_1$ decreases, the growing delocalized effects raise $\tilde{T}_2$ and lower $\tilde{T}_3$, and consequently increase the virtual temperature $T_v$.
This leads to the loss of coherence and  the improvement  of $\eta_g$  for low $E_1$.



Let us focus on the case of $\eta_g$  reaching its upper bound, shown by the solid red line in Fig. \ref{FigEtagMaxVsE1}.
For fixed the target and nonequilibrium fridge, i. e. $E_i$, $\gamma$, $T_2$ and $T_3$, we set $T_1=T_v$ and compare the ratio $\eta_g/\eta_c$ with the coherence of virtual qubit $C(\rho_v)$ in Fig. \ref{FigEtagMaxVsB3}.
One can find that, the internal coupling in the fridge extends the difference between $\eta_g$ and $\eta_c$, although the current $Q_{23}$ maintaining the nonequilibrium state of fridge is excluded in the definition of $\eta_g$.
On the other hand, the coherence $C(\rho_v)$ exhibits the opposite behavior of the COP.
This indicates that the quantumness prevents the fridge from reaching the Carnot performance.
The irreversibility comes from the fact that, to maintain the coherence, the two machine qubits do not reach thermodynamic equilibrium with their baths, although the heat current extracting energy from the target vanishes.
This offers a simple picture to understand the behavior of COP in \cite{PRE2013performance}, which approaches zero at the right endpoint of the cooling window.

From another point of view, the fridge can be regarded as two independent machine qubits, e. g. $\tilde{2}$ and $\tilde{3}$, running between two baths in temperatures $\tilde{T}_{2}$ and $\tilde{T}_{3}$.
When $T_1=T_v$, the  ratio of currents  extracted from qubit $1$  and $\tilde{3}$  $\tilde{\eta}_g=Q^g_1/\tilde{Q}^g_3=(\tilde{\beta}_2-\tilde{\beta}_3)/(\beta_1-\tilde{\beta}_2)$
reaches  the Carnot performance,
 where  $\tilde{Q}^g_{\nu}=- \tr [ \varepsilon_{\nu}\frac{\tilde{\sigma}_{\nu}^{z}}{2} \mathcal{D}_g(\rho_S) ]$  and $\tilde{\beta}_{\nu=2,3}=1/\tilde{T}_{\nu}$.
 However, in such picture, the two-qubit coherence is absent.
 One can  go further and  prove $Q^g_1+\tilde{Q}^g_{2}+\tilde{Q}^g_{3}=0$, $Q^g_2 = \tilde{Q}^g_{2} \cos^2\frac{\theta}{2}  +\tilde{Q}^g_{3}\sin^2\frac{\theta}{2} $ and $Q^g_3 = \tilde{Q}^g_{3} \cos^2\frac{\theta}{2}  +\tilde{Q}^g_{2}\sin^2\frac{\theta}{2} $.
 Then  the maximum COP is given by
  \begin{equation}\label{maxCOP}
\eta_{g} =\frac{\tilde{\beta}_2-\tilde{\beta}_3}{\beta_1 \cos \theta -\tilde{\beta}_2\cos^2\frac{\theta}{2}-\tilde{\beta}_3\sin^2\frac{\theta}{2}}.
\end{equation}
This view shows the relationships between the maximum COP and  the second law of thermodynamics.

\subsection{COP at maximum power}

\begin{figure}
 \begin{flushright}
\includegraphics[width=8.4cm]{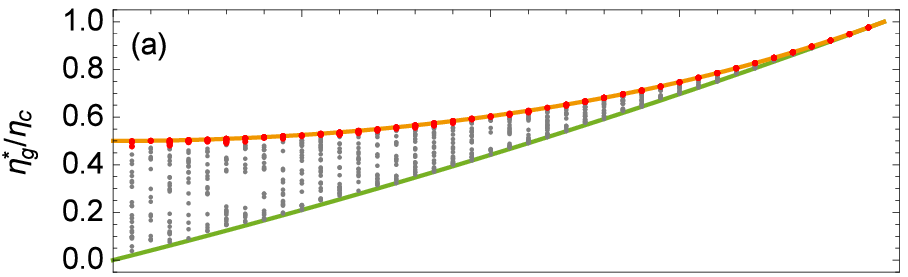}\\
\includegraphics[width=8.31cm]{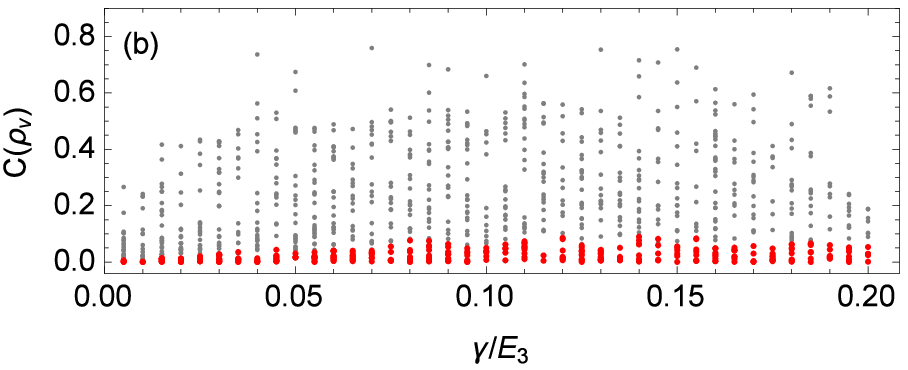}
\end{flushright}
\caption{(Color online)
Plots of 1000 random refrigerators with $\eta_c=1$ and $\gamma$ being random integral multiples of $E_3 \eta_c/200$ in (a) plane of $\eta^*_g/\eta_c$ vs $\gamma/E_3$ in company with the lines of $\eta^{*,\max}_g/\eta_c$ and $\eta^{*,\min}_g/\eta_c$, and in (b) plane of $C(\rho_v)$ vs $\gamma/E_3$, in which the red points satisfy $(\eta^{*,\max}_g-\eta^*_g)/(\eta^{*,\max}_g-\eta^{*,\min}_g)<0.05$.
} \label{FigEtaStarVsGamma}
\end{figure}

Another extreme is the COP at maximum power $\eta_g^*$,  i.e., when $Q_1^g$ is maximized, while the upper bound studied above requires $Q_1^g \to 0$.
This figure of merit is presented in  \cite{PRE2013performance},  and is  limited to some fractions of $\eta_c$ by many bounds for different setups \cite{SR2014quantum}.
Precisely, we maximize $Q_1^g$ by traversing the region of $E_1$ satisfying $Q_1^g>0$ when the other parameters are fixed, and substitute it in the definition of COP in Eq. (\ref{etag}).
We found that the amount of $\eta_g^*$ is tightly upper bounded by a function of $\gamma/E_3$ that
 \begin{equation}\label{etastar}
\eta_g^{*,\max}=\frac{\frac{1}{4} \eta_c^2 +4 \frac{\gamma^2}{E_3^2}}{\frac{1}{2}\eta_c - 2\frac{\gamma^2}{E_3^2}}.
\end{equation}
When $\gamma=0$, $\eta_g^{*,\max}=\frac{1}{2}\eta_c $, which returns the result of the original model obtained in \cite{PRE2013performance}.
It rises to $\eta_c$ as the value of $\gamma/E_3$ increases.
In Fig. \ref{FigEtaStarVsGamma} (a), we plot a set of random three-qubit models with a fixed $\eta_c$ to numerically verify the bound.
One  can also see a tightly lower bound $\eta_g^{*,\min} $, which is nothing but the minimum of $\eta_g$, corresponding to the lowest points of the curves with fixed $\gamma$ and $E_3$ in Fig. \ref{FigEtagMaxVsE1}.
Here a sufficient condition  for the upper bound $\eta_g^{*,\max}$ being saturated is the the high temperature limit, i. e. $E_1/T_1 \ll 1$ and  $E_{\mu}/T_{\nu} \ll 1$ with $\mu,\nu=2,3$.
One can start from the first-order approximation of $Q_1^g$ and analytically derive the expression in Eq. (\ref{etastar}).

A similar numerical analysis on these random models shows that, the upper bound of the thermodynamic COP at maximum power, $\eta_{tot}^*$, decreases from $\frac{1}{2}\eta_c$ to zero, as $\gamma/E_3$ increases from $0$ to $\eta_c/\sqrt{16+8 \eta_c}$.
And, the upper bound is approached by the same models as $\eta_g^{*,\max}$, which are marked in red in Fig. \ref{FigEtaStarVsGamma}.

We also plot the set of random models in the plane of  $C(\rho_v)$ vs $\gamma/E_3$ in Fig. \ref{FigEtaStarVsGamma} (b), to analyze the role of quantumness.
It is shown that, the nonequilibrium fridge with maximum power can have a larger virtual qubit coherence, e. g. the maximum $C(\rho_v)\simeq0.75$.
But the ones approaching $\eta_g^{*,\max}$  are obviously limited in a range of $C(\rho_v) \lesssim  0.1$.
These indicate that, the coherence of the virtual qubit holds back the COP at maximum power from reaching its upper bound, and therefore reflects irreversibility.

\section{CONCLUSIONS }\label{Conclu}

We study the three-qubit model of quantum absorption refrigerator with a strong coupling between the two machine qubits.
The thermal contact with two baths with a temperature difference makes the two-qubit machine in a nonequilibrium steady state with a quantum coherence between them.
The task is to cool  the target qubit, thermalized by a  cold bath, by coupling it with the machine.
The machine is refered to as a nonequilibrium fridge, as its performance characteristics  are determined by the two-qubit nonequilibrium state and its interaction with the target.
We define a COP only  taking  into account the currents caused by the arbitrarily weak tripartite interaction, and quantify the quantumness in the task of cooling by using the coherence of machine virtual qubit.

To act as a fridge, the internal coupling between the two machine qubits should be less than a critical value determined by their free Hamiltonian.
Such constraint is equivalent to the requirement that the virtual qubit coherence can be enhanced by the temperature difference between the two baths of the machine.
This result shows that the quantum coherence is beneficial to the nonequilibrium fridge.
However,  it is detrimental to the COP, although the heat current maintaining the coherence is excluded in the definition of COP.
The adverse effects are shown in two extreme cases, in one of which the heat current extracted  for the target approaches zero, and in the other the heat current is maximized.

In the representation of two free qubits in the diagnalized fridge, the delocalized dissipations on the total stationary state are equivalent to two local channels.
This provides an intuitive picture of  several results in our study, and consequently contributes to our understanding of delocalized effects on the three-qubit model with a strong tripartite interaction \cite{PRE2013performance,PRE2014re}.
Moreover, it is interesting to ask whether or when an interaction Hamiltonian can be regarded as a quantum machine running  among different degrees of freedom in a whole system.
In our model, the interaction $\mathcal{H}_g$ can reach the Carnot limit, when we consider it as a machine to cool the target by extracting free energy from qubit $\tilde{3}$ and exporting it into $\tilde{2}$ .

Our model can also serve as an example to verify the consistent of local approach with the second law of thermodynamics, which is questioned recently \cite{EPL2014local,EPL2016perturbative,PRA2007microscopic,ACMP2015nonequilibrium,OC2017two,gonzalez2017testing,hofer2017markovian}.
Here, we argue that the local approach is valid under the resonance between subsystems.
The arbitrarily weak interaction, allowing the subsystems to resonantly exchange energy, can be considered as the thermal contact among them \cite{PRE2012virtual}.
It is similar with the fact that only the effects of the resonant frequency are involved in the standard Lindblad master equation.

\begin{acknowledgments}
This work is supported by NSF of China (Grant No.11675119, No. 11575125, No.11105097).
\end{acknowledgments}

\bibliography{FrigeCoher}

\end{document}